\documentclass[conference]{IEEEtran}
\usepackage{blindtext, graphicx, tabularx, booktabs, amsmath, siunitx, xcolor, cite, color, soul, flushend, listings, siunitx, multirow}

\newcolumntype{R}[1]{>{\raggedright\arraybackslash}p{#1}}

\IEEEoverridecommandlockouts

\lstdefinestyle{customc}{
      breaklines=true,
      belowcaptionskip=0pt,
      abovecaptionskip=0pt,
      xleftmargin=3.5mm,
      xrightmargin=3.5mm,
      language=C,
      frame=single,
      showstringspaces=false,
      basicstyle=\footnotesize\ttfamily,
      keywordstyle=\bfseries\color{green!40!black},
      commentstyle=\itshape\color{purple!40!black},
      identifierstyle=\color{black},
      stringstyle=\color{orange},
}
    
\begin{document}

\title{
    NoX: a Compact Open-Source RISC-V Processor for  Multi-Processor Systems-on-Chip
}

 \author{
    \IEEEauthorblockN{
        Anderson I. Silva, Altamiro Susin, Fernanda L. Kastensmidt, Antonio Carlos S. Beck, Jose Rodrigo Azambuja
    }
  
    \IEEEauthorblockA{
        \textit{Federal University of Rio Grande do Sul (UFRGS) - Institute of Informatics - PPGC - PGMicro, Porto Alegre, Brazil}\\
        \{anderson.silva, altamiro.susin, fglima, caco, jose.azambuja\} @inf.ufrgs.br
    }   
}

\maketitle

\begin{abstract}
    IoT applications are one of the driving forces in making systems energy and power-efficient, given their resource constraints. However, because of security, latency, and transmission, we advocate for local computing through multi-processor systems-on-chip (MPSoCs) for edge computing. The RISC-V ISA has grown in academia and industry due to its flexibility. Still, available open-source cores cannot be seamlessly integrated into MPSoCs for a fast time to market. This paper presents NoX, a compact open-source plug-and-play 32-bit RISC-V core designed in System Verilog for efficient data processing in MPSoCs. NoX has a 4-stage single-issue in-order pipeline with full bypass, providing an efficient resource-constrained architecture. Compared to industry and academia resource-constrained RISC-V cores, NoX offers a better resource usage and performance trade-off.    
\end{abstract} 
    
\begin{IEEEkeywords}
    FreeRTOS, Muti-Processor System-on-Chip, Network-on-Chip, RISC-V
\end{IEEEkeywords}

\section{Introduction}
\label{sec:introduction}

    Internet of Things (IoT) devices are intelligent microprocessed components, often ubiquitous, that usually support cloud connectivity to interact with other devices \cite{iot_cloud}. Due to power and energy constraints, IoT devices often rely on resource-constrained microprocessors to control data acquisition and transmission, leaving the heavier processing to the cloud infrastructure. With an extensive number of deployed devices, latency becomes an issue \cite{iot_lat}. Also, such a cloud infrastructure needs to be evaluated as part of the solution. By offloading data processing, one must account for security issues \cite{iot_sec}, unpredictable transfer rates, and additional energy for sending large amounts of data over a congested network. Thus, we advocate for local processing with Multi-Processor System-on-Chip (MPSoC) architectures.

    MPSoCs consolidate multiple cores into a single platform to meet the growing demands for high performance. Additionally, application-specific hardware accelerators can be distributed over the network and more efficiently shared. Still, this rise in cores presents several challenges for traditional multicore platforms that rely on shared buses. These challenges brought to light issues like limited scalability and reduced communication efficiency due to constraints in bus bandwidth \cite{mpsoc_moraes}.

    To overcome these issues, Networks-on-Chip (NoC) have emerged as an essential component of MPSoC design due to their ability to provide efficient and versatile interconnects. Their modular and symmetric architecture supports various communication patterns, making it an attractive solution for multi-processor systems. Furthermore, NoCs offer better performance and scalability than traditional bus-based interconnects, which suffer from limited bandwidth and high latency as the number of processing elements increases. The primary concept of a NoC revolves around employing a hierarchical network structure with routers. This setup facilitates the smoother flow of packets between sender and receiver nodes while offering additional communication resources. By doing so, NoCs mitigate the energy and performance drawbacks associated with shared-bus communication, enabling multiple communication channels to operate concurrently \cite{mpsoc_wolf}.

    Along with optimized processing, heterogeneous MPSoC target architectures are important for energy efficiency \cite{mpsoc_spieck}, especially for IoT devices, where energy and power are often major system constraints \cite{edge_hmimz}. Each tile with its configuration can have an independent set of hardware accelerators with small microprocessors capable of handling small tasks and data management. In this configuration, the microprocessor coordinates the Control and Status Registers (CSRs) of the applications-specific hardware accelerators to perform their corresponding computation. As an independent grid, the MPSoC can power-gate or at least clock-gate independently tiles that are not part of the active data path of the current task.

    RISC-V is an ideal choice, considering that compact resource-constrained microprocessors are strong candidates for performing simple tasks and configuring and offloading data processing to a set of application-specific hardware accelerators. RISC-V is a versatile open-source Instruction Set Architecture (ISA) that supports a wide range of extensions, including vector processing, single and double floating-point precision, atomic instructions, and other features \cite{riscv_isa}. Due to their modularity, RISC-V processors can be tailored to given design constraints, such as resource usage. Therefore, one could design a custom microprocessor based on RISC-V with only the base subset of integer instructions, thus making a viable candidate for a NoC with resource-constrained microprocessors and multiple application-specific hardware accelerators for offloading local data processing \cite{hero_manycore_soc}.

    In this work, we propose NoX, a compact open-source 32-bit RISC-V core designed in System Verilog for efficient data processing in MPSoCs with FreeRTOS~\cite{FreeRTOS} support. NoX was designed for easy configuration and integration as part of a SoC, with a framework for standalone or interconnected simulation. It can easily interface with application-specific hardware accelerators. Compared to resource-constrained RISC-V cores in the Literature, NoX offers a plug-and-play processor with a better resource usage and performance trade-off.
    
\section{Background and Related Work}
\label{sec:related}

    RISC-V is an open-source ISA based on Reduced Instruction Set Computing (RISC) principles \cite{risc-v_0,risc-v_1,risc-v_2}. Unlike proprietary ISAs, such as ARM, x86, and MIPS, RISC-V is freely available for anyone to use, modify, and implement without needing to pay royalties or license fees, enabling an open market of open-source or proprietary processor cores. In recent years, many cores have been designed and made available by academia and industry \cite{ibex_0,cv32e40p,scr1,picorv,N22,E20,E21}, providing implementations for various applications domains, from resource-constrained to high-performance. Additionally, RISC-V has three base instruction sets and seven extensions, including instruction sets \textit{RV32I} (base 32-bit integer instruction set) and \textit{RV32E} (RV32I with a reduced register file to 16) and extensions \textit{M} (multiplication extension), \textit{C} (compressed instructions extension), \textit{A} (atomic extension), \textit{F}, (floating-point extension), \textit{V} (vector operations extension), and \textit{Zicsr} (Control and Status Register extension). 
    In the following, we briefly discuss seven resource-constrained RISC-V cores regarding performance (CoreMark/MHz) and area (kGE or LUTs).

    Ibex \cite{ibex_0} are area-optimized 2-stage 32-bit cores designed for control applications by ETH Zurich as part of the PULP Platform. They implement RV32E-C (micro) and RV32I-MC (small). Currently under development and maintenance under LowRISC \cite{lowrisc}, these cores can achieve up to 2.47 performance with an area of 26.6 kGE.

    The CV32E40P \cite{cv32e40p}, also known as RI5CY, is a 4-stage 32-bit core designed for a broader range of applications by ETH Zurich, also as part of the PULP Platform, that implements RV32I-MC. Additionally, it has an optional floating-point unit to support the F extension with DSP operations, hardware loops, SIMD, bit manipulation, and post-increment instructions. Currently under development and maintenance by the OpenHW group \cite{openhw}, it can achieve up to 3.19 performance at more than double the cost in Zero-riscy's area (40.7 kGE).

    SCR1\cite{scr1} is a 2-to-4-stage 32-bit open-source industry-grade silicon-proven core developed by Syntacore. It implements RV32I or RV32E. Additionally, it supports V, M, and C extensions through three predefined recommended configurations. It also includes full-wafer production and pre-build images for a wide range of FPGAs. Its performance ranges from 1.01 to 2.95 with an area up to 33 kGE.

    PicoRV32 \cite{picorv} is an open-source core designed by an individual targeting a small hardware footprint. It can be configured as RV32E or RV32I with C and M extensions. Designed for a 7-Series Xilinx FPGA, its area ranges from 750 to 2000 LUTs, achieving a 0.4 performance.
    
    N22\cite{N22} is a 2-stage 32-bit proprietary core by Andes Technologies that implements RV32I or RV32E with M, A, and C extensions. It also supports numerous hardware features, such as a configurable multiplier, branch predictor, and the AndeStar V5/V5e ISA, which still complies with RISC-V. Its performance reachs 3.97. However, the area is given in mm² and thus cannot be compared to other cores.

    E20 \cite{E20} and E21 \cite{E21} are cores designed by SiFive. E20 implements a 2-stage pipeline architecture with RV32I-MC. E21 expands E20 with a 3-stage pipeline and the A extension. Like N22, they provide area in mm², thus not comparable. However, their performance range from 2.51 to 3.1.

    NoX addresses some of these limitations while exploiting their advantages. For instance, its ready-to-use AMBA AXI and AHB interfaces seamlessly connect with various crossbars or interconnects, like PicoRV32, but with improved performance. Compared to SCR1, it supports FreeRTOS with similar performance metrics and smaller resource requirements.

\section{NoX Architecture}
\label{sec:nox_riscv}

    Even though heterogeneous MPSoCs, by definition, have different types of processing elements, they usually share similar computer architectures to ease software design. In this sense, a compact but fast processor is an attractive basic building block, which can also be used as a secondary processing tile. On top of this, more resources can be added to improve its performance, thus turning it into a primary processing tile capable of managing the MPSoC. Furthermore, the workload executed by secondary processing elements tends to be small compared to the workload running on the primary tile, as they do not need to split or merge the data being processed \cite{CAESAR-MPSoC}.

    Building on this compact yet fast processor feature, we leverage the RISC-V ISA to develop a compact open-source RISC-V processor that implements RV32I. The resulting core is referred to as the NoX RISC-V RV32I processor. It implements industry-standard interfaces such as AMBA AXI4 and AHBV5 for instruction fetching and load-store units, thus proving seamless integration.

    The NoX core has a 4-stage single-issue in-order pipeline with full bypassing to avoid data hazards that could result in delaying stalls. The only circumstances in which a stall may occur is if the core experiences back-pressure from the Load-and-Store Unit (LSU) or the Fetch stage as a result of an ongoing on-the-fly operation. This design follows a microarchitecture similar to that of the classic five-stage pipeline CPU \cite{riscv_arch_org_book}, except for the last two stages, LSU and Memory \& Writeback, which have been consolidated into a single stage.

    NoX currently does not have cache support. This decision is based on the class of applications foreseen. The application program will fit in the available program memory. At the same time, the data are transferred through the communication channel over the NoC. Another rationale for not implementing caches is to restrict the scope of development to reduce its complexity and to fit on the prototyping platform.

    The NoX processor design is freely available\footnote{GitHub repository: https://github.com/aignacio/nox} using the System Verilog IEEE 1800 standard \cite{ieee800}, with a few set parameters to configure its design. Basic macros can change the behavior of the design by selecting between \textit{synchronous} vs. \textit{asynchronous} and \textit{active-low} vs. \textit{active-high} reset. Additional configurations can also generate synchronous exceptions to trap errors and misaligned operations, which could also be performed by software in the LSU.

\begin{figure*}[!ht]
        \begin{center}
            \includegraphics[width=0.9\linewidth]{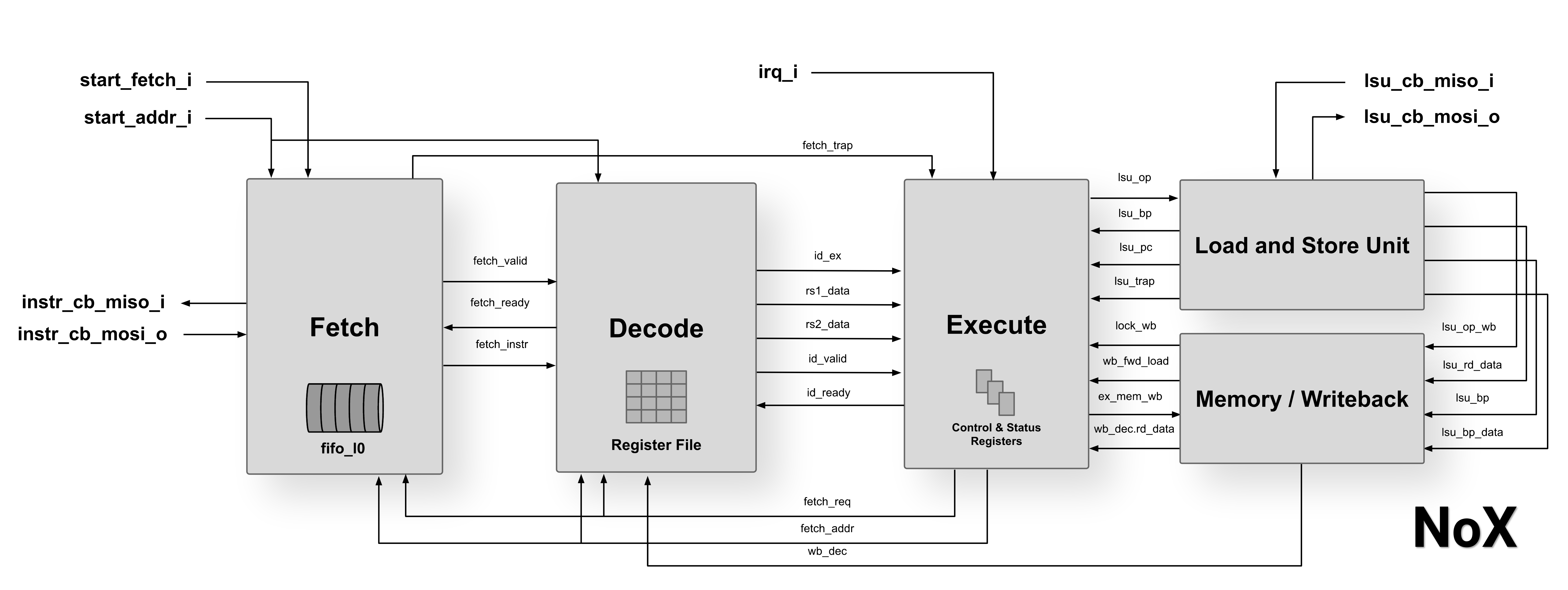}
        \end{center}        
        \caption{NoX's 4-stage pipeline microarchitecture}
        \label{fig:nox_cpu_diagram}
    \end{figure*}

    Fig. \ref{fig:nox_cpu_diagram} illustrates NoX's microarchitecture with its pipeline. In the following, we discuss each pipeline stage in detail.

    The \textbf{Fetch stage} features a configurable interface that supports two Advanced Microcontroller Bus Architecture (AMBA) interfaces, AXI and AHB. The utilization of AMBA interfaces aims to enable seamless core integration within a NoC interconnection infrastructure. This happens mainly because the \textit{fetch\_ready} signal removes the necessity for Tightly Coupled Memories (TCM), often seen in other resource-constrained RISC-V cores. By implementing support to fetch instructions from slower memories (\textit{i.e.}, with more than one clock cycle latency), NoX can alleviate memory preloading requirements. The result is a plug-and-play processing core that can easily fetch data from small buffer memories, like network interfaces, or even distributed ROM binaries connected directly to NoX's bus. Moreover, the Fetch stage has a configurable level-0 FIFO to optimize instruction pre-fetching for the following instruction execution.

    The \textbf{Decode stage} supports RISC-V's base 32-bit integer instruction set (RV32I) and implements the Zicsr extension, which includes manipulating internal CSRs, and machine mode, required to run FreeRTOS support. It interfaces directly with the Register File, simultaneously reading two registers and writing one register. Additionally, the decode stage manages various traps associated with decoding or stalling execution upon issuance of the Wait For Interrupt (WFI) instruction. This stage also implements different mechanisms to control memory data flow (for \textit{back-pressure} support from Fetch and LSU stages), including stall logic, which increments the area footprint of this stage, making it the largest one.

    The \textbf{Execute stage} executes 40 register-to-register and immediate operations and oversees access to the CSR address space with 6 additional instructions. Numerous core management operations are orchestrated within the CSR module, including handling various synchronous exceptions (traps) and asynchronous events (interrupts) and providing basic profiling registers for assessing processor performance. In this same stage, the machine mode privileged extension is managed with all sets of registers (\textit{e.g.}, mstatus, mepc, mie, etc.) being part of the same CSR block.

    The \textbf{Load and Store Unit (LSU) stage} can be implemented with configurable AXI or AHB interfaces, facilitating the dispatch of both aligned and unaligned transfers across these AMBA buses. Additionally, trap controls are integrated into this stage to address instances where the core detects bus errors or unexpected misalignments. Store instructions are executed and finished within this stage. However, load instructions still require processing in the Memory \& Writeback stage, where the corresponding destination register, located in the Register File, must be updated with data from the \textit{lsu\_rd\_data} signal.

    The \textbf{Memory \& Writeback stage} operates concurrently with the LSU stage. This design choice improves overall performance and resource usage, as most instructions use either the LSU or the Memory \& Writeback stage, thus reducing the number of clock cycles an instruction takes to transverse the pipeline, especially when stalls happen. 
    As a result, this parallelism minimizes the pipeline's commit latency without compromising the processor's performance.    
    
\section{Evaluation and Results}

    For this evaluation, we configure NoX with a 2-entry Fetch FIFO, AXI buses, and synchronous reset and compare it with the resource-constrained RISC-V cores discussed in Section \ref{sec:related}. We synthesized NoX for the FPGA flow using Xilinx's synthesis with Vivado 2022.1 targeting a Kintex-7 FPGA board (part XC7K325TFFG676-1) with a frequency target of 100 MHz to compare it with related works. For the ASIC flow, we synthesized NoX with YOSYS using Nangate 45 nm (FreePDK45) targeting a 250 MHz frequency.
    We evaluated NoX's performance based on the same configuration used to extract resource usage. To assess its performance accurately and make it fairly comparable to related works, we conducted the CoreMark benchmark~\cite{CoreMark}.    

    \subsection{Resource Usage}    

    In the context of MPSoC design, having a small processing core provides significant benefits. First, a smaller processing core design typically has lower complexity, reducing design time and development costs. Second, smaller designs usually consume less power, a crucial factor in embedded systems where power consumption is a significant concern. This can lead to reduced heat dissipation requirements and longer battery life in portable devices.

    Table \ref{tbl:nox_size} presents synthesis results for the FPGA (LUTS and Registers) and ASIC (kGE) flows and an area breakdown for the FPGA synthesis split into the five modules that compose the 4-stage pipeline. The Decode stage requires the most resources because it has an interlock mechanism for simultaneously handling bus stalls from the Fetch and the LSU stages. Additionally, the combinational logic used for decoding instructions, even though small in absolute numbers, is significant compared to the remaining resource-constrained stages. Finally, it has logic to interface with FreeRTOS (\textit{Zicsr} and machine-mode specific instructions), which could be removed to optimize resources. The Execute could also be optimized by removing registers and logic related to performance metrics.
    
    \begin{table}[!ht]
    \renewcommand{\arraystretch}{1.0}
    \caption{NoX breakdown area report.}
    \centering
    \label{tbl:nox_size}
    \resizebox{\columnwidth}{!}{
    \begin{tabular}{lrrr}
    \toprule
    \textbf{Pipeline Stage}               & \textbf{kGE} & \textbf{LUTs} & \textbf{Registers} \\ 
    \midrule
    \midrule
    Fetch (including I/O buffer)        & -       & 430                 & 143            \\
    Decode (including Register File)    & -       & 1,343               & 1,240          \\
    Execute (including CSR)             & -       & 253                 & 359            \\    
    Load and Store Unit         & -       & 611                 & 105            \\
    Memory \& Writeback                   & -       & 32                  & 33             \\
    \midrule
    \textbf{NoX}                         & \textbf{27.0}       & \textbf{2,665}               & \textbf{1,882}          \\
    \bottomrule
    \end{tabular}
    }
    \end{table}

    Table \ref{tbl:nox_area} compares our synthesis results to data collected from related works \cite{perf_ref}, including a NoX gain factor (smaller is better) that compares NoX directly to each related work. Note that we removed cores that do not offer mm² data (\textit{i.e.}, N22, E20, and E21), as many factors can affect the area, and we could not make a fair comparison to NoX. Considering the PULP platform, NoX has roughly the same requirements as the Ibex small (RV32I-MC) and less than RI5CY (35\% less). It is worth highlighting that the Ibex cores run their synthesis with a latch-based design, where flip-flops are replaced with latches, while we keep all flip-flops. This optimization reduces the number of resouce requirements but cannot be applied to FPGAs. Compared to SCR1, NoX has roughly the same requirements as RV32I-C and 20\% less than RV32I-MC. With regards to PicoRV32, NoX requires $2.8\times$ more area. For a better evaluation, we must take performance into account.
    
    \begin{table}[!ht]
    \renewcommand{\arraystretch}{1.0}
    \caption{NoX Area Comparison with Available Data.}
    \centering
    \label{tbl:nox_area}
    \resizebox{\columnwidth}{!}{
    \begin{tabular}{llrr}
    \toprule
    \multirow{2}{*}{\textbf{Processor Core}} & \multirow{2}{*}{\textbf{ISA}} & \multicolumn{1}{c}{\textbf{Resource Usage}} & \multirow{2}{*}{\textbf{NoX gain}} \\
                                             &                                     &\multicolumn{1}{c}{\textbf{kGE/LUTs}}&        \\
    \midrule
    \midrule
    \multirow{2}{*}{\textbf{NoX} (this work)} & \multirow{2}{*}{\textbf{RV32I-Zicsr}} 	    & \textbf{27.0 kGE}                  & \multirow{2}{*}{\textbf{1}}   \\
                                                                   &                    & \textbf{2,665 LUTs}          \\
    \midrule
    Ibex micro \cite{ibex_0,lowrisc}                        & RV32E-C		       & 16.9 kGE                  & 1.60                                 \\
    Ibex small \cite{ibex_0,lowrisc}                        & RV32I-MC	       & 26.6 kGE                  & 1.02                                 \\
    RI5CY \cite{cv32e40p,perf_ref}            & RV32I-MC		   & 40.7 kGE                  & 0.66                                 \\
    SCR1 \cite{scr1,perf_ref}                 & RV32E-C		       & 11.0 kGE                  & 2.45                                 \\
    SCR1 \cite{scr1,perf_ref}                 & RV32I-C		       & 26.0 kGE                  & 1.04                                 \\
    SCR1 \cite{scr1,perf_ref}                 & RV32I-MC		   & 33.0 kGE                  & 0.82                                 \\
    PicoRV32 \cite{picorv,perf_ref}           & RV32I-MC		   & 950 LUTs                  & 2.80                                 \\
    \bottomrule
    \end{tabular}
    }
    \end{table}

    \subsection{Performance}

    Table \ref{tbl:nox_perf} compares NoX's 2.5 CoreMark/MHz performance to data collected from related works \cite{perf_ref}, including a NoX gain factor (bigger is better) that compares NoX directly to each related work.
    NoX's performance surpassed the two smallest configurations proposed, with their micro configuration achieving 0.91 CoreMark/MHz and small configuration achieving 2.47 CoreMark/MHz. Despite lacking the M extension (Standard Extension for Integer Multiplication and Division), which is present in the small Ibex configuration and the SCR1 (RV32IC), the compact footprint of NoX achieves a favorable performance-to-area trade-off, making it advantageous for heterogeneous MPSoCs that necessitate a compact yet potent core for tile management duties.
    The improved performance of the NoX core can be attributed to its extensive use of forwarding across nearly all stages of its pipeline. This enables subsequent operations to bypass intermediate stages, thereby preventing core stalls. Additionally, the L0 pre-fetcher ensures a continuous instruction stream, guaranteeing an instruction available at each clock cycle. Consequently, even if an LSU operation experiences delays (e.g., due to bus arbitration), the core's single-issue front-end can concurrently fill the FIFO, maintaining efficient operation.

    \begin{table}[!ht]
    \renewcommand{\arraystretch}{1.0}
    \caption{NoX performance comparison.}
    \centering
    \label{tbl:nox_perf}
    \resizebox{\columnwidth}{!}{
    \begin{tabular}{llrr}
    \toprule
    \multirow{2}{*}{\textbf{Processor Core}} & \multirow{2}{*}{\textbf{ISA}} & \multicolumn{1}{c}{\textbf{Performance}} & \multirow{2}{*}{\textbf{NoX gain}} \\
                                             &                                     &\multicolumn{1}{c}{\textbf{CoreMark/MHz}}&        \\
    \midrule
    \midrule
    \textbf{NoX} (this work)                            & \textbf{RV32I-Zicsr} 		& \textbf{2.50}                  & \textbf{1}                             \\
    \midrule
    Ibex micro \cite{ibex_0,lowrisc}              & RV32E-C		& 0.90                  & 2.77                         \\
    Ibex small \cite{ibex_0,lowrisc}              & RV32I-MC		& 2.47                  & 1.01                         \\
    RI5CY \cite{cv32e40p,perf_ref}           & RV32I-MC		& 3.19                  & 0.78                         \\
    SCR1 \cite{scr1,perf_ref}                & RV32E-C		& 1.01                  & 2.48                         \\
    SCR1 \cite{scr1,perf_ref}                & RV32I-C		& 1.27                  & 1.97                         \\
    SCR1 \cite{scr1,perf_ref}                & RV32I-MC		& 2.95                  & 0.85                         \\
    PicoRV32 \cite{picorv,perf_ref}          & RV32I-MC		& 0.40                  & 6.25                         \\
    N22 \cite{N22,perf_ref}                  & RV32I-MAC	& 3.97                  & 0.63                         \\
    E20 \cite{E20,perf_ref}                  & RV32I-MC		& 2.51                  & 1.00                         \\
    E21 \cite{E21,perf_ref}                  & RV32I-MAC	& 3.10                  & 0.81                         \\
    \bottomrule
    \end{tabular}
    }
    \end{table}    

\section{Conclusions and Future Work}
\label{sec:conclusion}

    This paper presented NoX, a 32-bit compact open-source RISC-V processor with seamless integration to MPSoCs and FreeRTOS support. It was designed for ease of use through a platform that covers everything from high-level evaluation for SoC and MPSoC to FPGA and ASIC synthesis flows. Core connections are made via standardized bus and are configurable to meet specific application demands. Our evaluation shows that NoX performs better than RISC-V cores with the same resource requirements. In the future, we will incorporate NoX as an SoC into an MPSoC. Our final goal is to fabricate chiplets containing NoX-based MPSoCs.

\section{Acknowledgements}
\label{sec:acks}
    This work was financed in part by the Coordenação de Aperfeiçoamento de Pessoal de Nível Superior - Brasil (CAPES) - Finance Code 001, CNPq, and FAPERGS.

\ifCLASSOPTIONcaptionsoff
  \newpage
\fi

\bibliographystyle{IEEEtran}
\bibliography{ref}

\end{document}